\documentclass[twocolumn]{article}

\usepackage{graphicx}
\usepackage{mhchem}
\usepackage{eurosym}
\usepackage{color}

\title{Proof of concept of a zinc-silver battery for the extraction of
  energy from a concentration difference.}  

\author{M. Marino, L. Misuri, A. Carati, D. Brogioli}
\begin{document}

\maketitle

\begin{abstract}
The conversion of heat into current can be obtained by a process
with two stages. In the first one, the heat is used for distilling
a solution and obtaining two flows with different concentrations.
In the second stage, the two flows are sent to an electrochemical
cell that produces current by consuming the concentration
difference. In this paper, we propose such an electrochemical
cell, working with water solutions of zinc chloride. The cell
contains two electrodes, made respectively of zinc and silver
covered by silver chloride. The operation of the cell is analogous
to that of the capacitive mixing and of the ``mixing entropy battery'': the
electrodes are charged while dipped in the concentrated solution
and discharged when dipped in the diluted solution. The cyclic
operation allows us to extract a surplus of energy, at the expense
of the free energy of the concentration difference. We evaluate
the feasibility of such a cell for practical applications, and find
that a power up to 2~W per m$^2$ of surface of the electrodes can be
achieved.
\end{abstract}

\section{Introduction}

Salinity gradient power (SGP) is the production of renewable and clean
power from naturally available water reservoirs with different
salinity~\cite{pattle1954, norman1974, logan2012}, e.g. river and sea
water. SPG techniques can also be applied to solutions of salts,
different from sodium chloride, whose concentration difference is
produced by means of distillation~\cite{lamantia2011, carati2013}. In
the present paper, we focus on a zinc chloride solutions.

The two most mature SGP techniques are ``pressure-retarded
osmosis''~\cite{levenspiel1974, loeb1975, chung2012}, based on
semi-permeable membranes, and ``reverse
electrodialysis''~\cite{weinstein1976, post2007, post2008}, based on
ion-exchange membranes.  In order to avoid such expensive membranes, a
new family of electrochemical techniques have been recently
introduced: nanoporous materials are used as capacitive electrodes in
the ``capacitive mixing'' (CAPMIX) technique~\cite{brogioli2009,
  sales2010, brogioli2011, burheim2011, bijmans2012, liu2012,
  burheim2012, sales2012, rica2013}, while battery electrodes
undergoing redox reactions are used in the ``mixing entropy
battery''~\cite{lamantia2011,jia2013} and in the concentration cell
proposed in \cite{clampitt1976}. Since capacitors and batteries are
collectively called accumulators, this family of techniques will be
called ``accumulator mixing'' (AccMix). The device described in the
present paper (see Sect.~\ref{sect:cell:description}) belongs to this
family, being based on a zinc-silver chloride
battery~\cite{karpinski2000, crompton}; it is designed for working
with zinc chloride solutions.

\begin{figure}
% require cell_schemes.eps
\includegraphics{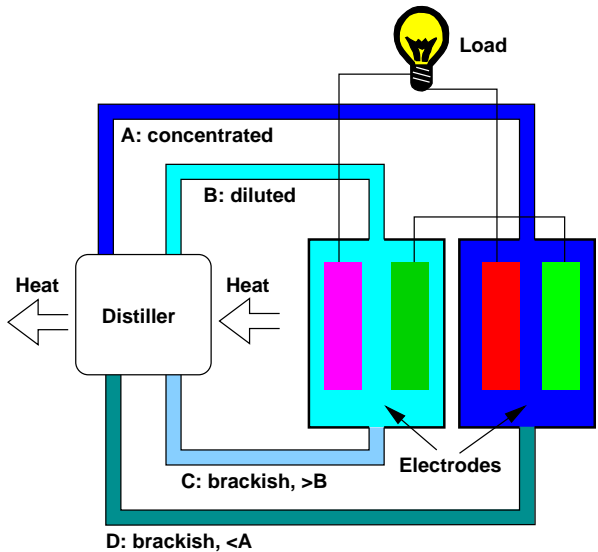}
\caption{{\bf Sketch of the heat-to-current converter system.} The electrochemical cells consume the
concentration difference between two flows A and B, using the available
free energy for producing an electrical current. The concentrations
are then restored by means of a distiller, that consumes heat. The system
is thus a heat-to-current converter.
\label{fig:sketch}}
\end{figure}

The application of SGP techniques to artificially produced
concentration differences has been limited, up to now, to the recovery
of by-products of other processes~\cite{tedesco2012, reapower}, or
even wastes such as \ce{CO2}~\cite{hamelers2013}. Recently, it has
also been proposed to use an SGP device in a closed cycle in which the
concentration difference is obtained by means of
distillation~\cite{lamantia2011, carati2013}, see
Fig.~\ref{fig:sketch}. 
% The heat is used to distill a water solution
% of the electrolyte, in such a way to generate two fluxes A and B of
% solutions at different concentrations. The concentration difference is
% then used in the second stage, i.e. the SGP device, in order to
% generate electrical current. When an electrical current flows through
% the electrodes of the cell, the two outgoing fluxes C and D have a
% reduced concentration difference. The operation of the device is in
% closed-cycle, i.e. the flows C and D are sent back to the
% distiller. 
The whole process is actually a heat-to-current
converter working in closed-cycle with respect to the
solutions. The first stage of the process, aimed at the regeneration
of the concentration difference, including the distiller, is not in
the scope of the present paper and was already discussed in
\cite{carati2013}.

In that paper, we have shown that the use of solutions with high
boiling point elevation leads to a high overall efficiency of the
energy conversion. For example, by making use of sodium hydroxide, the
efficiency of the global process was theoretically estimated, on
general thermodynamical grounds, to be of the order of 15\%. The
present paper focuses on another promising electrolyte, i.e. zinc
chloride. Indeed, in such a case on the one hand the solute presents a
high solubility and gives a high boiling point elevation (see
Sect.~\ref{sect:boiling:point}), on the other hand the conversion of
the salinity difference into electrical energy can be achieved by
means of the efficient zinc-silver AccMix cell that is described in
the present paper.

Another relevant feature of a heat-to-current current working with
zinc chloride solution can easily work with low-temperature heat
sources, i.e. in the range 80-130$^{\circ}$C, as will be shown in
Sect.~\ref{sect:cell:description}. So this technique can become a
promising one, with a wide range of applications, including solar
energy production with thermal collectors~\cite{kaushika2000,
  muschaweck2000, spirkl1998}, recovery of waste heat from industrial
processes, cogeneration of electrical current and heat for domestic
uses~\cite{rosato2013, pan2013}. Such techniques are particularly
appealing for small-scale applications and can become competitive with
the state-of-the art solutions, such as Stirling motors, organic
Rankine cycle turbines or thermoelectric elements, based on the
Seebeck effect~\cite{snyder2008, weidenkaff2008, frassie2013}. It's
worth noting that a two stage heat-to-current converter, involving
distillation of a ammonia/\ce{CO2} solution and a pressure-retarded
osmosis device has been proposed~\cite{mcginnis2007}.

A proof-of-concept of our zinc-silver AccMix cell is described in
Sect.~\ref{sect:cell:description}. The boiling point, and its relation
with the overall efficiency and the specific free energy of the
solution, is discussed in Sect.~\ref{sect:boiling:point}. The
equilibrium cell voltage is discussed in Sect.~\ref{sect:voltage}. The
discussion of the kinetics and the measurement of the power production
is presented in Sect.~\ref{sect:kinetics}.  In
Sect.~\ref{sect:silve_chloride} we report some unwanted effects,
leading to a reduction of the power production, which have been
observed in some cases, but which can be avoided by operating the cell
in a suitable regime.

\section{AccMix cell for zinc chloride}
\label{sect:cell:description}

In this section, we describe the implementation of an unusual 
SGP cell working with a zinc chloride solution, a good candidate for 
the closed-cycle heat-to-current converter based on SGP technology.
The cell is operated with AccMix cycles, and is inspired by the
zinc-silver chloride battery, that is known as one of the primary
batteries with the highest power density~\cite{karpinski2000,
crompton}. 

Figure~\ref{fig:sketch} shows a couple of AccMix cells, one for
each of the two flows at different concentrations. The voltages of
the two cells are different, due to the different chemical
potentials of the ions in the two solutions.  For this reason, a
current will flow through the load. This current cannot flow
indefinitely, because the electrodes in the cell at lower
concentration will be continuously consumed, and simultaneously
the ions will deposit on the electrodes in the higher
concentration cell. For this reason, the flows must be cyclically
exchanged.

\begin{figure}
% require CAPMIX_cycle.eps
\includegraphics{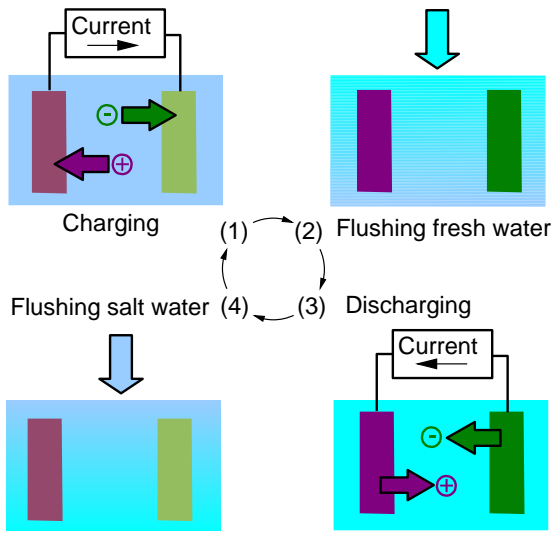}
\caption{{\bf Cycles of the AccMix technique.} The cycle starts with
  the cell filled with the high concentration solution. The phases
  are: (1) charge, (2) flow of dilute solution, (3) discharge, (4)
  flow of concentrated solution.
\label{fig:CAPMIX:cycle}
}
\end{figure}

Focusing on one of the cells,
the cycle is shown in Fig.~\ref{fig:CAPMIX:cycle}.
Each cell will thus undergo an AccMix cycle, analogous to the cycles
seen in CAPMIX and in the ``entropy mixing batteries''.

The energy comes from the voltage rise that takes place when the
solution concentration is changed, that is given by Nernst equation:
the charge is put into the cell at a lower voltage than that at which it is then
extracted. We have shown that the voltage is connected with the
adsorption of the solute during the step of the cycle in concentrated
solution, and its release in the diluted solution~\cite{rica2012_prl}:
the mixing is actually mediated by the temporary adsorption of the
ions inside the electrodes. The energy is extracted at the expense of
the concentration difference.

The AccMix cell is basically an accumulator, i.e. a battery or a
supercapacitor, because it must accumulate the charge during the step
A and give it back during the phase C. However, not all the battery
electrodes feature a voltage rise, i.e. a change of voltage upon
concentration changes, or, equivalently~\cite{rica2012_prl}, an
adsorption of ions during the charge or discharge step. For example,
in a usual lithium ion battery, the lithium ions are released by one
electrode and adsorbed by the other, both in charge and discharge
steps, and thus the current flow does not induce a net release or
adsorption of ions in the electrodes. For this reason, in the case of
the AccMix cell that has been proposed for lithium chloride solutions,
the manganese oxide (a typical positive electrode for lithium-ion
batteries) has been coupled to a silver / silver chloride electrode,
that interacts with chlorine ions~\cite{lamantia2011}. During the
charge, sodium and chlorine ions are adsorbed into the electrodes, and
they are later released during the discharge.

It's worth noting that the cell voltage, i.e. the potential difference
between the electrodes, has no relevance in AccMix
cycles. Indeed, there would be a technological advantage in having a
cell voltage close to zero, or even that the cell voltage changes
polarity upon the concentration change~\cite{brogioli2012}.

In order to find good electrodes for the zinc chloride solution, we
can profit from the wide literature about rechargeable batteries.
However, the requisites of our electrochemical cell are different from
the requisites of a usual rechargeable battery. The energy that is
extracted in each cycle is of the order of $Q\,\Delta V$, where $Q$ is
the charge that is exchanged in each cycle, and the voltage rise
$\Delta V$ is quite small (of the order of a fraction of one V). For
this reason, the losses must be extremely small. Some of the
requirements are:
\begin{enumerate}
\item The electrical charge efficiency of the charge/discharge cycles
  must be excellent: low self-discharge, i.e. low electrical charge
  leakage (it's an issue mainly for the supercapacitors electrodes
  based on activated carbon in the CDLE technique).
\item The energy efficiency of the charge/discharge cycles must be
  extremely high: low overvoltage.
\item Extremely high number of cycles.
\end{enumerate}

For reaching these parameters, we must use charge and discharge
currents that are much less than the currents usually applied in
rechargeable batteries, and apply them for much shorter times.
It is thus necessary to study the behaviour of the cells under
very unusual conditions, and such information is often lacking
in literature.

In the AccMix cell presented in this paper, one of the electrodes is a
zinc foil, and the other is a silver foil or felt, covered with silver
chloride. The electrolyte is a water solution of zinc chloride,
concentrated and diluted as necessary in steps A and C.

The interest in such an electrochemical cell stems from the fact
that a cell with similar electrodes is used as a primary battery,
with an excellent power density. A second reason of interest is
that zinc chloride has an extremely high solubility in water, and
leads to a very high boiling point elevation. This property is
connected with a high efficiency of the whole energy conversion
cycle~\cite{carati2013}.

The electrodes are 0.1~mm-thick 1$\times$1~cm foils (Alfa Aesar).  The
silver electrode is covered by silver chloride by means of
anodization, that is performed at 3~mA in a 3~M sodium chloride
solution for 1 hour. As an alternative, a silver wool, made of
100~$\mu$m wires, is employed; before the anodization, it is shaped as
a 1$\times$1~cm felt, around 1~mm thick.

The electrodes are kept at a distance of 0.6~mm, spaced by means of
two nylon wires, and clamped between two current collectors made of
graphite slabs 1~mm thick.

The solutions are obtained by dissolving zinc chloride (Alfa
Aesar) in deionized water. The pH of the solution is adjusted by
adding 1\% in mass of \ce{HCl}.

\begin{figure}
% require double_CAPMIX.eps
\includegraphics{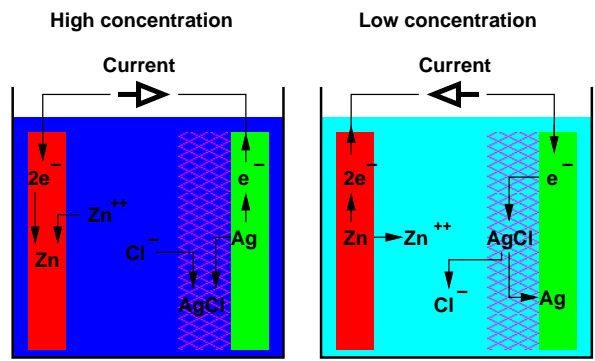}
\caption{{\bf AccMix cell for zinc chloride.} The reactions taking place
in the two steps of the AccMix cycle are shown. Legend:
\colorbox{blue}{\color{blue}{O}} Concentrated zinc chloride solution;
\colorbox{cyan}{\color{cyan}{O}} Dilute zinc chloride solution;
\colorbox{red}{\color{red}{O}} Zinc electrode;
\colorbox{cyan}{\color{magenta}{X}} Porous silver chloride;
\colorbox{green}{\color{green}{O}} Silver.
\label{fig:double:CAPMIX}}
\end{figure}

The reactions taking place in the cell during the AccMix cycle are
shown in Fig.~\ref{fig:double:CAPMIX}. The charging step (passive
phase) takes place when the cell is filled with the more
concentrated solution. The current flows from the zinc electrode
to the silver electrode. On the zinc electrode, zinc ions from the
solution are reduced to metallic zinc. The silver electrode is
oxidized, releasing silver ions that immediately precipitate
forming a silver chloride porous film on the surface of the
electrode. The discharging step (active phase) takes place when
the cell is filled with the less concentrated solution. The
current flows from the silver electrode to the zinc electrode. The
zinc electrode is oxidized, and the zinc ions are released into
the solution. The silver chloride on the surface of the silver
electrode is reduced, giving metallic silver and releasing
chloride ions into the solution.

It is evident that zinc chloride is temporarily stored into the
electrodes during the step in higher concentration, and is later
released into the less concentrated solution: we are actually
performing an accumulator-mediated mixing, i.e. an AccMix cycle. This
implies~\cite{rica2012_prl} that the cell, along the cycle, produces
energy.

The traditional scheme of the AccMix cell~\cite{brogioli2009,
lamantia2011} refers to a single cell, that is filled in different
phases with the two solutions at different concentration. In a
real working prototype, it will be useful to use a couple of
cells, performing cycles at 180$^{\circ}$, as shown in
Fig.~\ref{fig:sketch}.  In this way, one of the cells produces
energy, part of which is consumed by the other. However, since the
electrodes undergo accumulation or dissolution, the solution flows
must be periodically exchanged, in order to invert the
accumulation and dissolution: it is thus clear that the energy
production is actually obtained, also in this case, by an AccMix
cycle. In other words, the couple of cells is analogous to a
mechanical engine with two cylinders and pistons: the
active phase of one section of the engine drives the
passive phases of the other section, but the energy production
can be however evaluated by studying a single cylinder and
piston. Indeed, this is what we show in
Fig.~\ref{fig:double:CAPMIX}: in the present paper, we make experiments on a
single cell.

\section{Boiling point elevation}
\label{sect:boiling:point}

\begin{figure}
% require boilingpoint.txt
\includegraphics{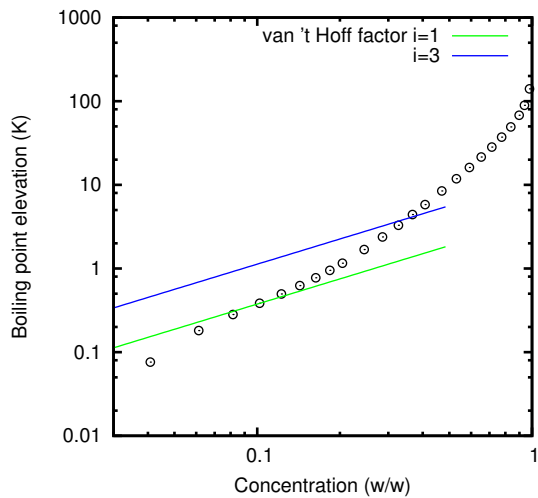}
\caption{{\bf Boiling point elevation of the \ce{ZnCl2} solutions at
    1~atm pressure.} Data provided by Italshell~\cite{italschell}. The
  concentration is expressed as the ratio between the mass of the
  solute and the total mass of the solution. The lines are the linear
  approximation for dilute solutions, for two different values of the
  van 't Hoff factor. The actual boiling point elevation has an
  evident deviation from the ideal solution, also at low
  concentrations. A quite high boiling temperature is observed in
  concentrated solutions.
\label{fig:boiling:point}}
\end{figure}

The choice of the solution, i.e. zinc chloride in water, is based on
the evaluation of the overall efficiency of the conversion of heat
into current. Indeed, as already recalled, in \cite{carati2013} it was
shown that, in order to increase the efficiency, a solute with a high
boiling point elevation should be preferred~\cite{carati2013}. The
elevation of the boiling temperature of solutions of zinc chloride in
water is reported in Fig.~\ref{fig:boiling:point} (data provided by
Italshell~\cite{italschell}). In this graph, and throughout this
paper, the concentration is reported as the ratio between the mass of
zinc chloride with respect to the total mass of the solution. The
boiling point elevation that can be obtained is quite high, and thus
this solution is promising. For ideal solutions, the boiling point
elevation is expected to be $K_b b_{solute} i$, where $K_b$ is the
ebullioscopic constant (0.512~K~Kg/mol) for water, $b_{solute}$ is the
molality of the solute, and $i$ is the van 't Hoff factor. This
formula predicts a linear dependence of the boiling point elevation at
low concentrations. Instead, the actual boiling point elevation
increases much more than linearly. This effect can be explained by
assuming activity coefficients significantly different from 1, and
considering that the zinc chloride forms complexes.

Working with solutions at about 0\% and 80\%, we expect a theoretical
efficiency of 10\% with a single-effect distiller
\cite{carati2013}. Working at atmospheric pressure, the solution
evaporates at 140$^{\circ}$C and condenses at 100$^{\circ}$C. However,
the pressure can be reduced, as usual in distillation plants, so that
the two temperatures become respectively around 80 and
45$^{\circ}$C. An actual thermic device, reaching the above-mentioned
efficiency, has been designed by Italschell~\cite{italschell}.

\begin{figure}
% require energia.txt
\includegraphics{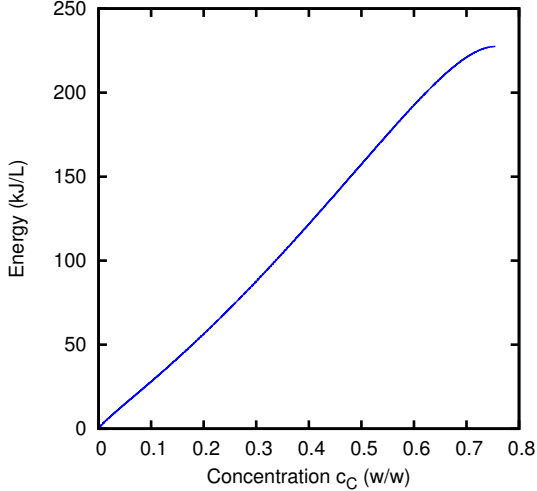}
\caption{{\bf Available free energy.} The graph shows the energy that
  can be ideally produced by a cell per liter of solvent in the
  diluted solution, as a function of the concentration of the outgoing
  diluted solution. The concentrated solution has a mass fraction
  $C_A=75.5$\%, considered constant during the cycle, $c_D=c_A$. The
  step of the cycle in dilute solution starts with the cell filled
  with fresh water, $c_B=0$, and stops at a concentration $c_C$.
\label{fig:energia:libera}}
\end{figure}

The boiling temperature as a function of the concentration allows
us to evaluate the free energy of the solution~\cite{carati2013}.
The free energy that is lost by the solutions, during the AccMix
cycle, is converted into electrical current. An ideal way to operate
the cell, for which the efficiency of this conversion could reach the
maximum theoretical value, is the following.  The flow rate of the
concentrated solution, at concentration $c_A$, is so high that its
concentration does not change significantly during its passage
through the cell, so that $c_D\approx c_A$.  The other flow from
the distiller is pure water, $c_B=0$, and the solution exits from
the cell at concentration $c_C$. The reduction of free energy,
corresponding to the electrical energy that could 
be extracted by means of such an
ideal cycle per liter of solvent in the diluted solution,  
is shown in Fig.~\ref{fig:energia:libera} as a
function of $c_C$, for $c_A=c_D=75.5$\%.

\section{Cell voltage and voltage rise}
\label{sect:voltage}

\begin{figure}
% require potenziali.txt
\includegraphics{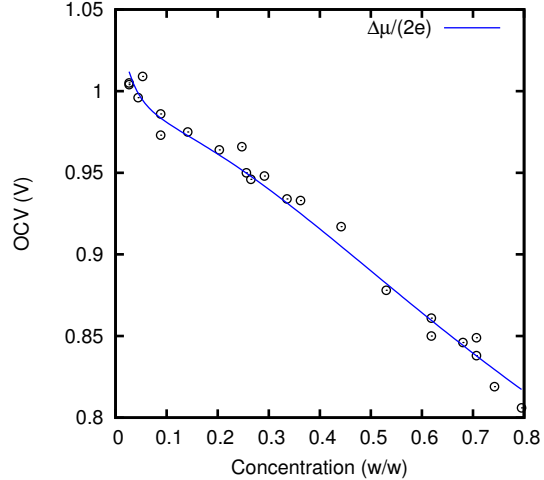}
\caption{{\bf Cell voltage as a function of the concentration.} The
  voltage is measured in open circuit. The solid line has been obtained from
  the chemical potential, in turn calculated from the boiling point
  elevation.
\label{fig:potenziale}}
\end{figure}

We measured the cell voltage in open circuit. The result is shown
in Fig.~\ref{fig:potenziale}.

The solid line is calculated from the boiling point elevation
curve. This curve actually allows us to evaluate the derivative of 
the Gibbs free energy
with respect to the number of units of \ce{ZnCl2} that are dissolved in the
solution~\cite{carati2013}. This quantity represents an ``overall''
chemical potential $\mu$ of the solute.\footnote{This quantity is defined as
  ``overall'' because it does not refer to an actual chemical species
  present in the solution. Indeed, the solute is actually present in
  various forms: ions \ce{Zn++} and \ce{Cl-}, undissociated molecule
  \ce{ZnCl2}, and various complexes, e.g. \ce{ZnCl3-}.}  The passage
of a pair of electrons through the cell leads to the adsorption or
release of a unit of \ce{ZnCl2}. Assuming that the cell works
reversibly, we expect to observe a cell voltage $E= \Delta\mu/(2e)$,
where $\Delta \mu$ is the difference between the chemical potential in
the solution and that (assumed to be constant) on the electrodes in
the solid state. This fact
can be rigorously shown in terms of Nernst equation; indeed, the
``overall'' chemical potential $\mu$ corresponds at equilibrium also 
to the chemical
potential of the undissociated molecule \ce{ZnCl2} and to the sum of
the chemical potentials $\mu_{\ce{Zn++}}+2\mu_{\ce{Cl-}}$ of the
dissociated ions \ce{Zn++} and \ce{Cl-}, also if the dissociation is
not complete, or complexes such as \ce{ZnCl3-}\cite{jorne1982} are
formed.

It must be noticed that the boiling point elevation allows us to evaluate
only the variation of the chemical potential $\mu$ with respect to the
concentration; the values are thus adjusted with a suitable additive constant.

In Fig.~\ref{fig:potenziale}, it can be noticed that the open circuit
cell voltage is indeed very close to $\Delta\mu/(2e)$. This shows that, at
least in quasi-stationary conditions with a low current, the AccMix
cell can work close to reversibility, i.e.~it can in principle extract
an energy equal to the overall reduction of the free energy of the solutions.
The voltage rise, that is one of the main parameters of the AccMix
cell, is the difference between the cell voltages at the
concentrations of the feed solutions.

It can be noticed that the cell voltage is nearly linear in the
concentration, and quite different from the logarithmic dependence
that would be expected by approximating the activities with the
concentrations. However, this is not surprising: at the highest
concentrations that we consider the solution is mainly
composed by zinc chloride, and water is the minor component.

\section{Electrokinetics measurements and AccMix cycles}
\label{sect:kinetics}

\begin{figure}
% require cs10.txt
\includegraphics{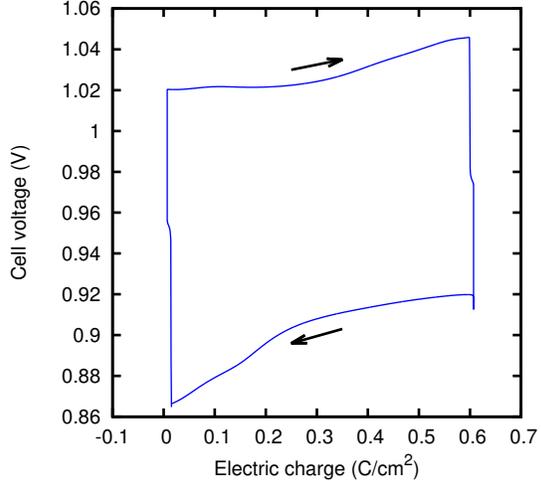}
\caption{{\bf Example of charge-discharge cycle.} The cycle is clockwise.}
\label{ciclo_cs}
\end{figure}

We characterized the kinetic behaviour of the cell by means of
charge/discharge cycles performed at constant concentration, at
various values of the current. The cycles were composed by a charge
phase at a given positive constant current for 60~s, followed by a
discharge phase at the opposite current for 60~s. The cycle duration
was chosen in order to mirror the actual duration of the typical
AccMix cycles. The dissipated power is evaluated as the average of
the product of the current and voltage over five cycles. The cycles can
be effectively represented by plotting the cell voltage as function of
the electric charge which has flown through the cell (i.e.~the integral
of the current with respect to time). An example of such a plot is given
in Fig.~\ref{ciclo_cs}. The energy dissipated during a cycle is equal
to the area enclosed by the path.

\begin{figure}
% require dati30.txt
% require dati50b.txt
% require dati80f.txt
% require dati05.txt
\includegraphics{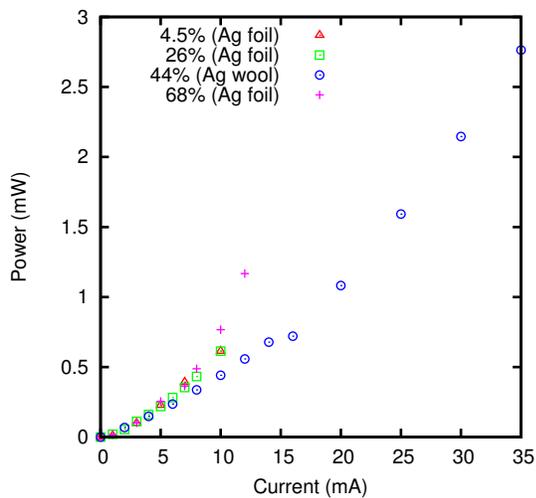}
\caption{{\bf Average dissipated power during charge-discharge cycles
    in \ce{ZnCl2} solutions.} The various data sets refer to different
  concentrations and different electrodes. The size of the electrodes
  is 1$\times$1~cm; in the case of the the \ce{Ag} wool, this is the
  macroscopic dimension of the felt. }
\label{fig:power:cycles}
\end{figure}

The measured power dissipation is shown in
Fig.~\ref{fig:power:cycles}. It can be noticed that the power
dissipation is slightly higher at 68\% concentration. This can be
explained by considering that the resistivity of the \ce{ZnCl2}
solution has a minimum of 0.1~$\Omega$m at 3.7~M
concentration~\cite{thomas1982}, and it increases at higher
concentrations, e.g. it is 1.3~$\Omega$m at 80\%.

The power dissipation is lower in the case of the felt obtained with
the silver wool, in agreement with the larger effective surface: that
is approximately 15~cm$^2$ for the 1$\times$1~cm felts we used.

\begin{figure}
% require dati30.txt
% require dati50b.txt
% require dati80g.txt
% require dati05.txt
\includegraphics{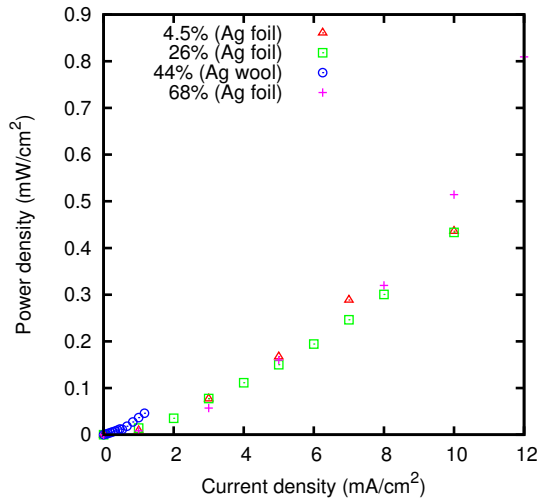}
\caption{{\bf Average dissipated power density at AgCl/Ag electrode.}
\label{fig:dissipated:ag}}
\end{figure}

By means of a reference electrode, we tried to roughly measure the
potentials of the single electrodes of the cell, 
in order to evaluate their individual
contributions to power dissipation (the
contribution of the resistance of the solution, although still
visible in the case of the 68\% solution, is relatively
small).

The power dissipation at the \ce{Ag}/\ce{AgCl} electrode is reported
in Fig.~\ref{fig:dissipated:ag}, expressed in terms of current per
unit surface. It can be noticed that the power dissipation in the
case of the silver wool is higher; this means that the geometric
surface of the wires is not completely available for the reaction. It
is possible that the silver chloride film, that forms during the
anodization, insulates a part of the microwires forming the silver
wool with respect to the others, and with respect to the current
collector.

\begin{figure}
% require dati30.txt
% require dati50b.txt
% require dati05.txt
% require dati80g.txt
\includegraphics{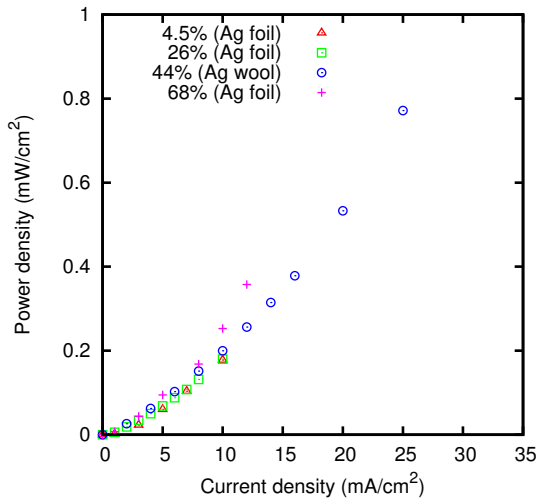}
\caption{{\bf Average dissipated power density at Zn electrode.}
\label{fig:dissipated:zn}}
\end{figure}

The power dissipation at the zinc electrode is reported in
Fig.~\ref{fig:dissipated:zn}. It can be noticed that the dissipation
on the zinc electrode is less than the dissipation on the silver
electrode, which thus represents the current bottleneck of the
technique.

\begin{figure}
% require ac4.txt
\includegraphics{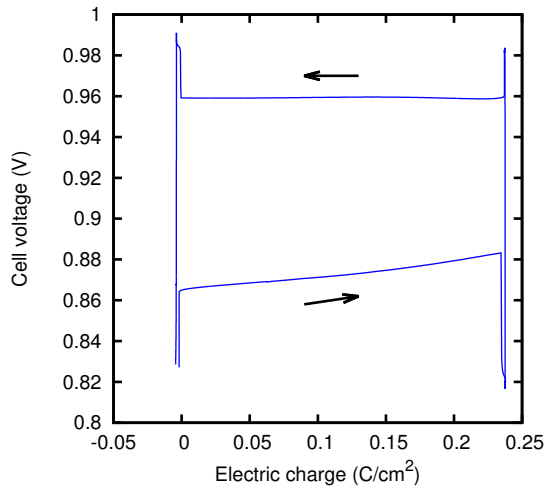}
\caption{{\bf Example of AccMix cycle.} The cycle is
  counterclockwise. The voltage rise is approximately 160~mV. The
  overvoltage, that reduces the cycle's surface, results from the
  flowing current of 4~mA.}
\label{ciclo_accmix}
\end{figure}

With our experimental setup we performed a series of AccMix cycles, 
which were composed
of a charge phase of 60 s in concentrated solution at a given current,
and a discharge phase of 60 s in diluted solution at the opposite
current. Between any two consecutive phases of this type (the
``active'' phases of the cycle) there was an interval of
60 s at 0 current, during which we could transfer the
electrodes from the container of the concentrated solution to that of
the diluted solution or vice-versa. A plot of such a series of cycles in the
$(Q,V)$ plane is shown in Fig.~\ref{ciclo_accmix}. At variance with
the charge-discharge cycles of Fig.~\ref{ciclo_cs}, here the cycles are
counterclockwise, and the enclosed area represents an energy which is
delivered by the cell to the external circuit. The average produced
power was calculated as the ratio between the average produced energy
per cycle and the sum of the durations of the two active phases of the 
cycle (120 s in our case), therefore neglecting the intervals required 
for the transfer of the electrodes. 

\begin{figure}
% require capmix_cycles.txt
\includegraphics{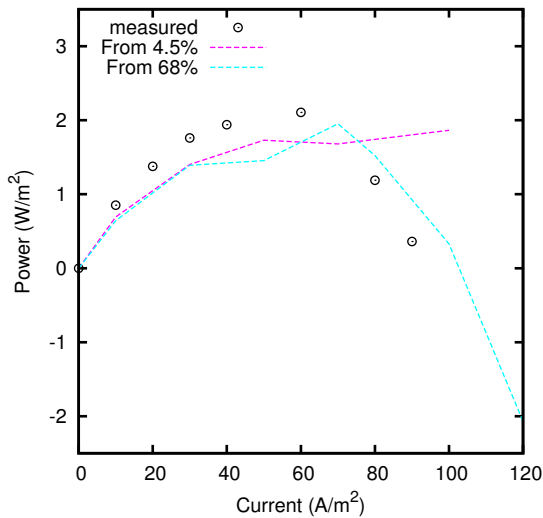}
\caption{{\bf Average gained power in AccMix cycles}. The
  concentrations of the \ce{ZnCl2} solutions are $c_A=4.5$\% and
  $c_C=68$\%. The lines represent an evaluation of the power
  production obtained by subtracting the power dissipation, measured
  during charge/discharge cycles, from an ``ideal'' power production,
  obtained by assuming a voltage rise of 160~mV and no overvoltage.}
\label{fig:accmix}
\end{figure}

We measured the power production at various current intensities. 
The results are shown in Fig.~\ref{fig:accmix}. A maximum of approximately 
2~W per square meter of electrode is obtained.

\section{Silver chloride film}
\label{sect:silve_chloride}

\begin{figure}
% require cicli_carica.txt
\includegraphics{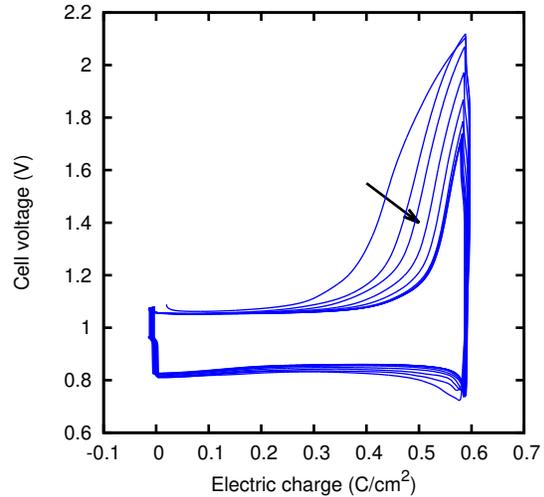}
\caption{{\bf Charge-discharge cycles with voltage increase at the end
    of the charge phase.} The cycles are clockwise. The arrow shows
  the reduction of the peak with time. }
\label{carica}
\end{figure}

\begin{figure}
% require cicli_scarica.txt
\includegraphics{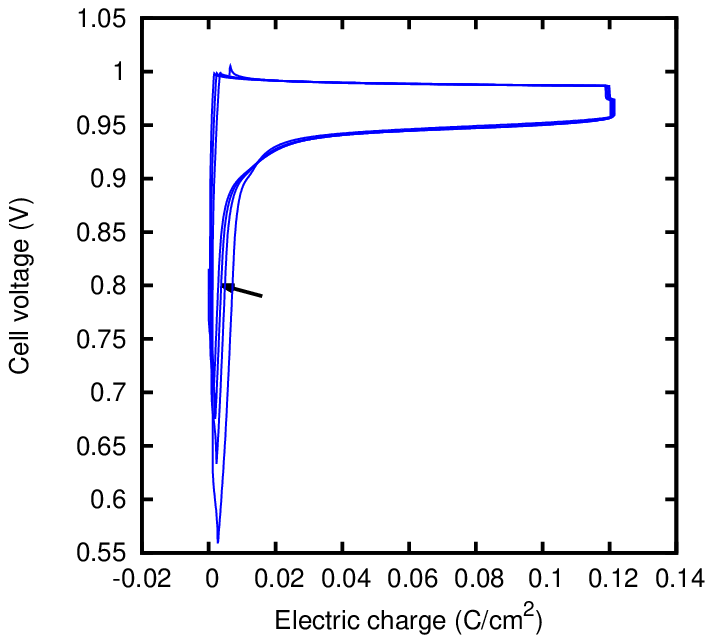}
\caption{{\bf Charge-discharge cycles with voltage fall at the end of
    the discharge phase.} The cycles are clockwise. A strong deviation
  from the rectangular shape of the cycle is evident at the end of the
  discharge phase (the branch at lower voltage). }
\label{scarica}
\end{figure}

The behavior of the electrodes apparently keeps memory of their recent
history. This was seen in some particular cases of charge--discharge 
cycles as those described in Sect.~\ref{sect:kinetics}.
In Figs.~\ref{carica}-\ref{scarica} we show two series of such cycles 
which were performed under slightly different conditions.
In the first case, we see that the potential increases near the end of
the charging phase, and that such an increase occurs closer and closer
to the end of the phase, as cycles are repeated. In the second case we
see instead that a potential fall occurs closer and closer to the end
of the discharge phase.  A possible interpretation of such a behavior
is that during the charging phase silver chloride grows more easily at
sites from which it has just been removed in the preceding cycles, and
similarly that during the discharge phase silver chloride is more
easily removed from sites in which it has just been grown in the
preceding cycles.  It is however difficult to systematically reproduce
these phenomena. The best performance is of course obtained when
either the potential increase in the charge or the potential fall in
the discharge are avoided, and the cycle is represented by a roughly regular
parallelogram in the $(Q, V)$ plane. In this case in fact the area of the
plane enclosed by the cycle, which represents a dissipated power, is
minimized.

% \section{Possible evolutions and alternative electrolytes}

% A possible problem that could take place is that a small amount of
% silver chloride actually dissolves. The solution becomes likely
% saturated with a very small quantity of silver ions, that could
% deposit on the zinc electrodes. It is however possible to reduce this
% problem by

% Possibili problemi. Una piccola parte dell'argento va in soluzione e
% puo' depositarsi sulle lastre di zinco. Bisogna pensare ad apposite
% trappole, e.g. una spugna di ferro da sostituire periodicamente.

% Ridurre il costo: Ag/Pb/Cu, Cl/Br (confrontare solubilita')

% \begin{table}
% \begin{tabular}{cccc}
%  & \ce{Ag+} & \ce{Cu+}\\
% \ce{Cl-} & 1.77$\times$10$^{-10}$M$^2$ & 1.72$\times$10$^{-7}$M$^2$\\
% \ce{Br-} & 5.2$\times$10$^{-13}$M$^2$  & 6.3$\times$10$^{-9}$M$^2$
% \end{tabular}
% \caption{{\bf Solubility product of some salts.}}
% \end{table}

% In order to reduce the solubility of the silver salt, it is possible
% to substitute the chlorine ions with bromine ions.

% In order to reduce the cost of the device, \ce{Cu} could be used for
% the positive-potential-rise electrode. In this case, it would be
% better to use bromide ions; however, the solubility of the resulting
% salt (\ce{CuBr}) would be higher than the solubility of \ce{AgCl}.

\section{Conclusions}

We have shown that our AccMix cell is efficient in converting the free
energy of the zinc chloride solution into electrical current.
The fact that the open-circuit voltage practically coincides with
the value theoretically expected from thermodynamical considerations
(Fig.~\ref{fig:potenziale}) means that the
efficiency actually approaches unity as the current goes to
zero. The maximum power production we achieved is around 2~W per square
meter of electrode. By using microporous materials, e.g. sintered
powders, we can have a surface of the order of 100~cm$^2$/g; this
leads to a cost of the order of ten euros per W. This is a relatively
good result: for comparison, reverse electrodialysis applied to sea
and river water gives a power output of the order of 1~W per square
meter of couple of membranes, having a cost of the order of 100 euro
each.

As already noticed, the bottleneck of the system is the silver/silver
chloride electrode. For this reason, further improvements might be
obtained by replacing the silver electrode with other materials
interacting with chlorine ions, e.g.~conducting polymers like
polypirrole.

\section*{Acknowledgments}
We thank Alessandro Atti, Pino Gherardi and Paolo Turroni~\cite{et}
for the collaboration in the development of the concept of the
closed-cycle heat-to-current converter.  We thank Carlo Guidi,
Francesco Gugole, Paolo Cesana and Duilio Calura
(Italschell~\cite{italschell}) for the project on the
concentrator/distiller for the zinc chloride.  We thank Luigi Galgani
and Maarten Biesheuvel for useful discussions and suggestions on the
manuscript.  The research leading to these results received funding
from the European Union Seventh Framework Programme (FP7/2007-2013)
under agreement no. 256868, CAPMIX project.

\bibliographystyle{plain}
\bibliography{zinco}

\begin{thebibliography}{10}

\bibitem{italschell}
Italschell, Monza, Italy, http://www.italschell.it/.

\bibitem{et}
ET-EcoinnovativeTechnologies S.r.l, spin-off ENEA (National agency for new
  technologies, energy and sustainable economic development, Bologna, Italy).

\bibitem{reapower}
Reapower project, under {FP7} ({FP}7/2007-2013).

\bibitem{bijmans2012}
M.~F.~M. Bijmans, O.~S. Burheim, M.~Bryjak, A.~Delgado, P.~Hack, F.~Mantegazza,
  S.~Tennisson, and H.~V.~M. Hamelers.
\newblock Capmix- deploying capacitors for salt gradient power extraction.
\newblock {\em Energy Procedia}, 20:108--115, 2012.

\bibitem{brogioli2009}
D.~Brogioli.
\newblock Extracting renewable energy from a salinity difference using a
  capacitor.
\newblock {\em Phys. Rev. Lett.}, 103:058501, 2009.

\bibitem{brogioli2012}
D.~Brogioli et~al.
\newblock Exploiting the spontaneous potential of the electrodes used in
  capacitive mixing technique for extraction of salinity-difference energy.
\newblock {\em Energ. Environ. Sci.}, 5(12):9870--9880, 2012.

\bibitem{brogioli2011}
D.~Brogioli, R.~Zhao, and P.~M. Biesheuvel.
\newblock A prototype cell for extracting energy from a water salinity
  difference by means of double layer expansion in nanoporous carbon
  electrodes.
\newblock {\em Energ. Environ. Sci.}, 4:772--777, 2011.

\bibitem{burheim2011}
O.~Burheim, B.B. Sales, O.~Schaezle, F.~Liu, and H.~V.~M. Hamelers.
\newblock Auto generative capacitive mixing of sea and river water by the use
  of membranes.
\newblock ASME 2011 International Mechanical Engineering Congress and
  Exposition (IMECE2011) (Denver, Colorado, USA) p. 483--492, paper no.
  IMECE2011-63459, 2011.

\bibitem{burheim2012}
O.~S. Burheim, F.~Liu, B.~B. Sales, O.~Schaetzle, C.~J.~N. Buisman, and H.V.~M.
  Hamelers.
\newblock Faster time response by the use of wire electrodes in capacitive
  salinity gradient energy systems.
\newblock {\em J. Phys. Chem. C}, 116:19203--19210, 2012.

\bibitem{carati2013}
A.~Carati, M.~Marino, and D.~Brogioli.
\newblock Theoretical thermodynamic analysis of a closed-cycle process for the
  conversion of heat into electrical energy by means of a distiller and an
  electrochemical cell.
\newblock http://arxiv.org/abs/1309.3643.

\bibitem{chung2012}
T.~S. Chung, X.~Li, R.~C. Ong, Q.~Ge~H. Wang, and G.~Han.
\newblock Emerging forward osmosis (fo) technologies and challenges ahead for
  clean water and clean energy applications.
\newblock {\em Curr. Opin. Chem. Eng.}, 1(3):246--257, 2012.

\bibitem{clampitt1976}
B.~H. Clampitt and F.~E. Kiviat.
\newblock Energy recovery from saline water by means of electrochemical cells.
\newblock {\em Science}, 194:719--720, 1976.

\bibitem{crompton}
T.~P.~J. Crompton.
\newblock {\em Battery Reference Book}.
\newblock Newnes, third edition edition, 2000.

\bibitem{frassie2013}
G.~Fraisse, J.~Ramousse, D.~Sgorlon, and C.~Goupil.
\newblock Comparison of different modeling approaches for thermoelectric elemen
  ts.
\newblock {\em Energ. Convers. Manag.}, 65(SI):351--356, 2013.

\bibitem{hamelers2013}
H.~V.~M. Hamelers, O.~Schaetzle, P.~M.~Biesheuvel J.~M.~{Paz-Garc\'ia}, and
  C.~J.~N. Buisman.
\newblock Harvesting energy from co$_2$ emissions.
\newblock {\em Environ. Sci. Technol. Lett.}, 1:31--35, 2014.

\bibitem{jia2013}
Z.~Jia, B.~Wang, S.~Song, and Y.~Fan.
\newblock A membrane-less na ion battery-based capmix cell for energy
  extraction using water salinity gradients.
\newblock {\em RSC Adv.}, 3:26205--26209, 2013.

\bibitem{jorne1982}
J.~Jorne and W.~T. Ho.
\newblock Transference numbers of zinc in zinc-chloride battery electrolytes.
\newblock {\em J. Electrochem. Soc.}, 129(5):907--912, 1982.

\bibitem{karpinski2000}
A.~P. Karpinski, S.~J. Russell, J.~R. Serenyi, and J.~P. Murphy.
\newblock Silver based batteries for high power applications.
\newblock {\em J. Power Sources}, 91(1):77--82, 2000.

\bibitem{kaushika2000}
N.~D. Kaushika and K.~S. Reddy.
\newblock Performance of a low cost solar paraboloidal dish steam generating
  system.
\newblock {\em Energ. Convers. Manag.}, 41(7):713--726, 2000.

\bibitem{lamantia2011}
F.~{L}a Mantia, M.~Pasta, H.~D. Deshazer, B.~E. Logan, and Y.~Cui.
\newblock Batteries for efficient energy extraction from a water salinity
  difference.
\newblock {\em Nano Lett.}, 11:1810--1813, 2011.

\bibitem{levenspiel1974}
O.~Levenspiel and N.~de~Vevers.
\newblock The osmotic pump.
\newblock {\em Science}, 183:157--160, 1974.

\bibitem{liu2012}
F.~Liu, O.~Schaetzle, B.~B. Sales, M.~Saakes, C.~J.~N. Buisman, and H.~V.~M.
  Hamelers.
\newblock Effect of additional charging and current density on the performance
  of capacitive energy extraction based on donnan potential.
\newblock {\em Energy Environ. Sci.}, 5:8642--8650, 2012.

\bibitem{loeb1975}
S.~Loeb.
\newblock Osmotic power plant.
\newblock {\em Science}, 189:654--655, 1975.

\bibitem{logan2012}
B.~E. Logan and M.~Elimelech.
\newblock Membrane-based processes for sustainable power generation using
  water.
\newblock {\em Nature}, 488(7411):313--319, 2012.

\bibitem{tedesco2012}
Tedesco M., Cipollina A., Tamburini A., van Baak~W., and Micale G.
\newblock Modelling the reverse electrodialysis process withseawater and
  concentrated brines.
\newblock {\em Desalination and Water Treatment}, 49:404--424, 2012.

\bibitem{mcginnis2007}
R.~L. McGinnis, J.~R. McCutcheon, and M.~Elimelech.
\newblock A novel ammonia-carbon dioxide osmotic heat engine for power
  generation.
\newblock {\em J. Membr. Sci.}, 305(1--2):13--19, 2007.

\bibitem{muschaweck2000}
J.~Muschaweck, W.~Spirkl, A.~Timinger, N.~Benz, M.~Dorfler, M.~Gut, and
  E.~Kose.
\newblock Optimized reflectors for non-tracking solar collectors with tubular
  absorbers.
\newblock {\em Solar energy}, 68(2):151--159, 2000.

\bibitem{norman1974}
R.~W. Norman.
\newblock Water salination: a source of energy.
\newblock {\em Science}, 186:350--353, 1974.

\bibitem{pan2013}
T.~H. Pan, D.~L. Xu, A.~M. Li, S.~S. Shieh, and S.~S. Jang.
\newblock Efficiency improvement of cogeneration system using statistical
  model.
\newblock {\em Energ. Convers. Manag.}, 68:169--176, 2013.

\bibitem{pattle1954}
R.~E. Pattle.
\newblock Production of electric power by mixing fresh and salt water in the
  hydroelectric pile.
\newblock {\em Nature}, 174:660, 1954.

\bibitem{post2008}
J.~W. Post, H.~V.~M. Hamelers, and C.~J.~N. Buisman.
\newblock {\em Env. Sci. Techn.}, 42:5785--5790, 2008.

\bibitem{post2007}
J.~W. Post, J.~Veerman, H.~V.~M. Hamelers, G.~J.~W. Euverink, S.~J. Metz,
  K.~Nymeijer, and C.~J.~N. Buisman.
\newblock Salinity-gradient power: Evaluation of pressure-retarded osmosis and
  reverse electrodialysis.
\newblock {\em J. Membrane Sci.}, 288:218--230, 2007.

\bibitem{rica2012_prl}
R.~A. Rica, R.~Ziano, D.~Salerno, F.~Mantegazza, and D.~Brogioli.
\newblock Thermodynamic relation between voltage-concentration dependence and
  salt adsorption in electrochemical cells.
\newblock {\em Phys. Rev. Lett.}, 109:156103, 2012.

\bibitem{rica2013}
R.~A. Rica, R.~Ziano, D.~Salerno, F.~Mantegazza, R.~{van Roij}, and
  D.~Brogioli.
\newblock Capacitive mixing for harvesting the free energy of solutions at
  different concentrations.
\newblock {\em Entropy}, 15(4):1388--1407, 2013.

\bibitem{rosato2013}
A.~Rosato and S.~Sibilio.
\newblock Performance assessment of a micro-cogeneration system under realistic
  operating conditions.
\newblock {\em Energ. Convers. Manag.}, 70:149--162, 2013.

\bibitem{sales2012}
B.~B. Sales, O.~S. Burheim, F.~Liu, O.~Schaetzle, C.~J.~N. Buisman, and
  H.~V.~M. Hamelers.
\newblock Impact of wire geometry in energy extraction from salinity
  differences using capacitive technology.
\newblock {\em Environ. Sci. Technol.}, 46(21):12203--12208, 2012.

\bibitem{sales2010}
B.~B. Sales et~al.
\newblock Direct power production from a water salinity difference in a
  membrane-modified supercapacitor flow cell.
\newblock {\em Env. Sci. Techn.}, 44:5661, 2010.

\bibitem{snyder2008}
G.~J. Snyder and E.~S Toberer.
\newblock Complex thermoelectric materials.
\newblock {\em Nat. Mater.}, 7(2):105--114, 2008.

\bibitem{spirkl1998}
W.~Spirkl, H.~Ries, J.~Muschaweck, and R.~Winston.
\newblock Nontracking solar concentrators.
\newblock {\em Solar energy}, 62(2):113--120, 1998.

\bibitem{thomas1982}
B.~K. Thomas and D.~J. Fray.
\newblock The conductivity of aqueous zinc chloride solutions.
\newblock {\em Journal of Applied Electrochemistry}, 12(1):1--5, 1982.

\bibitem{weidenkaff2008}
A.~Weidenkaff, R.~Robert, A.~M. Aguirre, L.~Bocher, T.~Lippert, and
  S.~Canulescu.
\newblock Development of thermoelectric oxides for renewable energy conversion
  technologies.
\newblock {\em Renew. Energ.}, 33(2):342--347, 2008.

\bibitem{weinstein1976}
J.~N. Weinstein and F.~B. Leitz.
\newblock Electric power from differences in salinity: the dyalitic battery.
\newblock {\em Science}, 191:557--559, 1976.

\end{thebibliography}

\end{document}